\documentclass[%
 reprint,
superscriptaddress,
 amsmath,amssymb,
 aps,
prb,
]{revtex4-2}

\usepackage{graphicx}
\usepackage{dcolumn}
\usepackage{bm}
\usepackage{siunitx}

\begin{document}

\preprint{APS/123-QED}

\title{Anisotropic Second Harmonic Generation From Monocrystalline Gold Flakes}

\author{Sergejs Boroviks}
\affiliation{Centre for Nano Optics, University of Southern Denmark, Campusvej 55, DK-5230
Odense M, Denmark}

\author{Torgom Yezekyan}
\affiliation{Centre for Nano Optics, University of Southern Denmark, Campusvej 55, DK-5230
Odense M, Denmark}

\author{\'{A}lvaro Rodríguez Echarri}
\affiliation{ICFO --- Institut de Ciencies Fotoniques, The Barcelona Institute of Science and Technology, 08860 Castelldefels (Barcelona), Spain}
\affiliation{Centre for Nano Optics, University of Southern Denmark, Campusvej 55, DK-5230
Odense M, Denmark}

\author{F.~Javier~Garc\'{i}a~de~Abajo}
\affiliation{ICFO --- Institut de Ciencies Fotoniques, The Barcelona Institute of Science and Technology, 08860 Castelldefels (Barcelona), Spain}
\affiliation{ICREA --- Instituci\'o Catalana de Recerca i Estudis Avan\c{c}ats, Passeig~Llu\'{\i}s~Companys~23, 08010 Barcelona, Spain}

\author{Joel D. Cox}
\affiliation{Centre for Nano Optics, University of Southern Denmark, Campusvej 55, DK-5230
Odense M, Denmark}
\affiliation{Danish Institute for Advanced Study, University of Southern Denmark, Campusvej 55, DK-5230 Odense M, Denmark}

\author{Sergey I. Bozhevolnyi}%
\affiliation{Centre for Nano Optics, University of Southern Denmark, Campusvej 55, DK-5230
Odense M, Denmark}
\affiliation{Danish Institute for Advanced Study, University of Southern Denmark, Campusvej 55, DK-5230 Odense M, Denmark}

\author{N. Asger Mortensen}%
\affiliation{Centre for Nano Optics, University of Southern Denmark, Campusvej 55, DK-5230
Odense M, Denmark}
\affiliation{Danish Institute for Advanced Study, University of Southern Denmark, Campusvej 55, DK-5230 Odense M, Denmark}

\author{Christian Wolff}
\affiliation{Centre for Nano Optics, University of Southern Denmark, Campusvej 55, DK-5230
Odense M, Denmark}
\email{cwo@mci.sdu.dk}

\date{\today}

\begin{abstract}
Noble metals with well-defined crystallographic orientation constitute an appealing class of materials for controlling light-matter interactions on the nanoscale. Nonlinear optical processes, being particularly sensitive to anisotropy, are a natural and versatile probe of crystallinity in nano-optical devices. Here we study the nonlinear optical response of monocrystalline gold flakes, revealing a polarization dependence in second-harmonic generation from the \{111\} surface that is markedly absent in polycrystalline films. Apart from suggesting an approach for directional enhancement of nonlinear response in plasmonic systems, we anticipate that our findings can be used as a rapid and non-destructive method for characterization of crystal quality and orientation that may be of significant importance in future applications. 
\end{abstract}

\maketitle

The experimental discovery of second-harmonic generation (SHG) from a metallic boundary~\cite{PRL1965Brown} triggered extensive studies of nonlinear optical processes in metals that have spanned several decades. One of the early pioneering works~\cite{PhysRev1968Bloembergen} reported SHG in reflection from various metal surfaces and established the central theoretical treatment of this phenomenon. Effectively, the induced second-order polarization can be expressed as~\cite{NonlinearOptics2019Boyd}
\begin{equation}
    P_i (2\omega) = \sum_{jk} \chi_{ijk}^{(2)} E_j(\omega) E_k(\omega), 
\end{equation}
where $\chi_{ijk}^{(2)}$ is the effective second-order susceptibility tensor component, $i$, $j$, $k$ indices run over Cartesian coordinates $x$, $y$, $z$, and $E_i$ denotes the electric field vector component $i$ at the fundamental frequency $\omega$.

In their natural form, noble-metal crystals occur in centro-symmetric form, as face-centre-cubic (FCC) lattice crystals, rendering SHG symmetry-forbidden within the dipole approximation, while multipole interactions enable weak second-order nonlinear effects even in centro-symmetric media~\cite{PRB1988Guyot}. Nonetheless, SHG from noble metals can be observed in reflection from metal-dielectric interfaces, where the inversion symmetry of the bulk crystal is broken and the second-order nonlinear response is no longer prohibited. The exact nature of this phenomenon remained a matter of debate over decades in the nonlinear optics community, with the main focus placed on the separation between the dipole surface-like contribution (occurring due to breaking of translation symmetry at the interface) and quadrupole bulk-like contributions (associated with strong field gradients at the metal surface)~\cite{PRB1980Sipe, PRB1986Guyot,PRB1987Weber, PRB1988Schaich}. Important theoretical developments by Rudnick and Stern~\cite{PRB1971Rudnick} eventually showed that both surface and bulk induced currents may be important, and the two can be related to the corresponding $\chi^{(2)}$ tensor via phenomenological constants. A significant breakthrough in the experimental investigation of this question was made by Wang {\it et al.}~\cite{PRB2009Wang}, who showed, using sophisticated two-beam SHG measurements, that surface dipole effects dominate the nonlinearity in sputtered (implying polycrystalline) gold films. 

In contrast to surfaces of polycrystalline metal films, their monocrystalline counterparts are known to exhibit anisotropic surface SHG, which manifests in the dependence of the emitted light polarization on the orientation of the sample relative to the excitation polarization. One of the first experimental observations was obtained from the silicon \{111\} surface~\cite{PRL1983Tom}, later also extended to surfaces of noble-metal crystals~\cite{CPL1989Friedrich, PRB1990Lupke, PRB1993Koos, JChemPhys1993Wong, JPhysChemB1998Yagi, JChemPhys2001Matranga}. These experimental demonstrations agree well with predictions of the phenomenological model developed by Sipe {\it et al.}~\cite{PRB1987Sipe}, who derived expressions for the second-order susceptibility tensor components arising from both bulk- and surface-like contributions, based on symmetry considerations for three conventional FCC crystal surfaces --- \{100\}, \{110\}, and \{111\}.

In this work, we experimentally investigate SHG from monocrystalline gold flakes, which have recently become a material platform of interest for high-quality plasmonic devices~\cite{NatComm2010Huang, Nanophotonics2020Kaltenecker, Nanophotonics2020Siampour}, owing to the moderate optical damping associated with both their atomic-scale surface flatness and high degree of crystallinity~\cite{OMEX2017Mejard, OMEX2019Boroviks}. We show that the \{111\} surface of the gold flakes exhibits polarization-dependent second-harmonic generation, which is nearly order of magnitude stronger than that observed in isotropic SH emission of evaporated gold films. The anisotropic nonlinear optical response we report here may be of crucial importance for directional enhancement of harmonic generation in metal nanostructures~\cite{Science2005Muhlschlegel, OptExp2011Renger, NatPhotonics2012Kauranen, ACSNano2015Butet}.
Furthermore, we suggest that our findings can be used as a fast and non-destructive method for characterization of crystallinity and lattice orientation of metal substrates, which is an important asset for ongoing experimental activities on quantum surface effects and future plasmonic applications~\cite{PRL2019Grossmann,Echarri:2020,ACSNano2019El-Fattah}.

\begin{figure}[!ht]
    \centering\includegraphics[width = 0.9\linewidth]{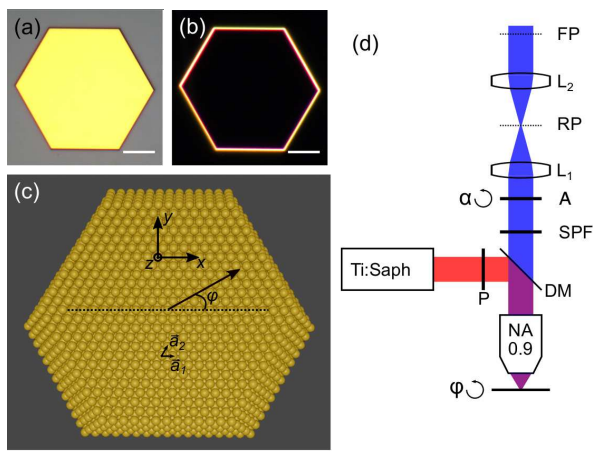}
    \caption{\label{fig1} (a) Bright-field and (b) dark-field optical microscope images of the investigated gold flake. Scale bars are 20\,\si{\micro \meter}. (c) Illustration of the crystal lattice of the gold flake with indicated \{111\} surface lattice primitive vectors $\vec{a}_1$ and $\vec{a}_2$, the sample orientation angle $\phi$ with respect to excitation polarization and coordinate axis. (d) Schematics of the experimental setup: a laser beam from a Ti:Saph oscillator is filtered using a polarizer (P) and then focused on to a sample with an apochromatic objective. The nonlinear reflection is collected with the same objective and filtered out using a dichroic mirror (DM) and a short-pass filter (SPF). The polarization of the nonlinear signal is resolved using an analyzer (A), whereas the signal can be recorded either in the real image plane (RP) or Fourier image plane (FP) created by lenses L$_1$ and L$_2$, respectively.}
\end{figure}

Monocrystalline gold flake samples are synthesized using a modified Brust--Schiffrin method~\cite{JChemSoc1994Brust}, as described in detail elsewhere~\cite{OMEX2018Boroviks}. The gold flakes have high aspect ratio, typically spanning $\approx100$\,\si{\micro \meter} in lateral directions and a few hundred nanometers in thickness. Optical micrographs of the gold flake sample investigated in this Letter are presented in Fig.~\ref{fig1}(a) and (b), revealing the absence of defects or contamination on the surface of the sample. The gold flakes have a FCC crystal lattice and their hexagonal faces are of the \{111\} type~\cite{CrystGrowthDes2018Krauss}, as illustrated schematically in Fig.~\ref{fig1}(c) alongside our definition of the coordinate frame and relevant angles. The evaporated gold film sample was prepared in a Cryofox TORNADO 405 evaporation system by Polyteknik, by depositing a \SI{120}{nm} gold layer on a silicon substrate previously coated with a \SI{3}{nm} titanium layer for adhesion purposes. Such gold films are known to be intrinsically polycrystalline~\cite{ACSPhotonics2015McPeak}.

The experimental setup, schematically depicted in Fig.~\ref{fig1} (d), is a custom-made scanning nonlinear microscope equipped with interchangeable detectors, namely: spectrograph (Ocean Optics QEPro), photo-multiplying tube (Hammatsu R3235-01) and sCMOS camera (Thorlabs CS235MU). The sample is mounted on a computer-controlled scanning stage with an in-plane rotation stage, which allows us to investigate the anisotropy of the sample. The laser source used in the experiment is a mode-locked Titanium-Sapphire (Ti:Saph) laser (Tsunami 3941 by Spectra-Physics), tunable in the 780--840\,\si{nm} wavelength range, with pulse duration of $\lessapprox1$\,\si{ps}. The excitation and collection of the SH signal is performed using a high (0.9) numerical aperture (NA) objective (Olympus MPLFN100X). All measurements described in this work were performed with $\approx100$\,\si{mW} average power focused to the diffraction-limited ($\approx450$\,\si{nm}) excitation spot. Finally, the polarization state and angular distribution of the SH emission can be analyzed, for example, by recording polarization-resolved Fourier images. 

\begin{figure}[!ht]
    \centering\includegraphics[width = 0.78\linewidth]{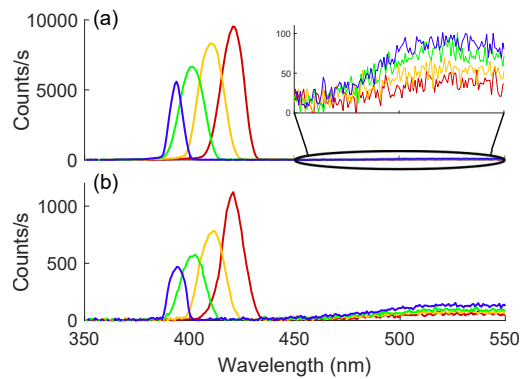}
    \caption{\label{fig2} Spectrum of the nonlinear reflection from (a) a \{111\} surface of a monocrystalline gold flake and (b) evaporated polycrystalline gold. The inset in (a) shows a zoom-in of the measured TPL spectral region (450--550\,\si{nm}). Curve colors indicate excitation wavelength: 780\,\si{nm} (blue), 800\,\si{nm} (green), 820\,\si{nm} (yellow), and 840\,\si{nm} (red).}
\end{figure}

First, we study spectra of the nonlinear emission from the monocrystalline gold flake in the 780--840\,\si{nm} range of excitation wavelengths and compare it with the evaporated gold film. Fig.~\ref{fig2} shows that the SH intensity generated at a crystalline surface is almost one order of magnitude stronger than at its polycrystalline counterpart, which can be explained by the presence of a large anisotropic contribution in former and absence of it in latter, as will be discussed below.

The main trend for both monocrystalline and polycrystalline surfaces is that the efficiency of SHG increases with increasing wavelength, which may be attributed to the excitation of a plasmon resonance at the fundamental frequency~\cite{JoAP2004Krause,Sensors2014Wang} and reduced coupling of SH fields to the interband transitions~\cite{JChemPhys1993Wong}. In addition, two-photon luminescence (TPL), characterized by emission in the broad 450--550\,\si{nm} range, is found to decrease with increasing wavelength. TPL arises from a third-order nonlinear process~\cite{PRB2009Biagioni}, which is incoherent and known to be significantly enhanced at rough surfaces and higher excitation energies~\cite{PRB1986Boyd,PRL2003Bozhevolnyi}. Here, we also observe approximately 35\% less TPL emission from the smoother crystalline gold surface.

The symmetry group of the large \{111\} surfaces of our gold flakes is not the hexagonal point group $\mathcal{C}_{6v}$ that one might expect from the arrangement of the top layer of atoms; it is rather the trigonal point group $\mathcal{C}_{3v}$ due to the stacking order of the FCC crystal breaking some of the hexagonal symmetries. Based on the above mentioned symmetry argument, it was generally shown that the second-order susceptibility $\chi_{ijk}^{(2)}$, which is a third-rank tensor, should have 11 non-zero and 5 independent elements~\cite{PRB1987Sipe,JEPT1986Aktsipetrov}. One can separate anisotropic and isotropic contributions, as some of the tensor components explicitly reflect the sample orientation angle $\phi$ with respect to the excitation polarization: $\chi^{(2)}_{xxx} = -\chi^{(2)}_{xyy}  = -\chi^{(2)}_{yxy}$ are associated with SHG $\propto \sin (3\phi)$; $\chi^{(2)}_{xxy} =\chi^{(2)}_{yxx} = -\chi^{(2)}_{yyy}$ are associated with SHG $\propto \cos (3\phi)$; and $\chi^{(2)}_{zxx} = \chi^{(2)}_{zyy}$, $\chi^{(2)}_{xzx}=\chi^{(2)}_{yzz}$ and $\chi^{(2)}_{zzz}$ are associated with isotropic SHG. The indices of the $\chi^{(2)}$ tensor correspond to the axis defined in Fig.~\ref{fig1}(c). Noticeably, anisotropic tensor components that give rise to in-plane induced polarization interact only with the in-plane $E$-field components $E_x$ and $E_y$.

With this phenomenological treatment, the dependence of SHG on the sample orientation angle can be described with a simple model for co-polarized and cross-polarized detection: 
\begin{subequations}
    \begin{align}
        I_{\parallel}(2\omega) \propto & | a_{i\parallel} -  a_{a\parallel} \sin 3 \phi |^2 I^2(\omega),\\
        I_{\perp}(2\omega) \propto & |a_{i\perp} + a_{a\perp} \cos 3 \phi |^2 I^2(\omega),
    \end{align}
\end{subequations}
where $I_{\parallel}(2\omega)$ and $I_{\perp}(2\omega)$ denote co-polarized and cross-polarized detected intensity, $I(\omega)$ is the excitation intensity, while $a_i$ and $a_a$ are the corresponding complex-valued phenomenological constants describing isotropic and anisotropic contributions to SHG, respectively.

\begin{figure}[!ht]
    \centering\includegraphics[width =\linewidth]{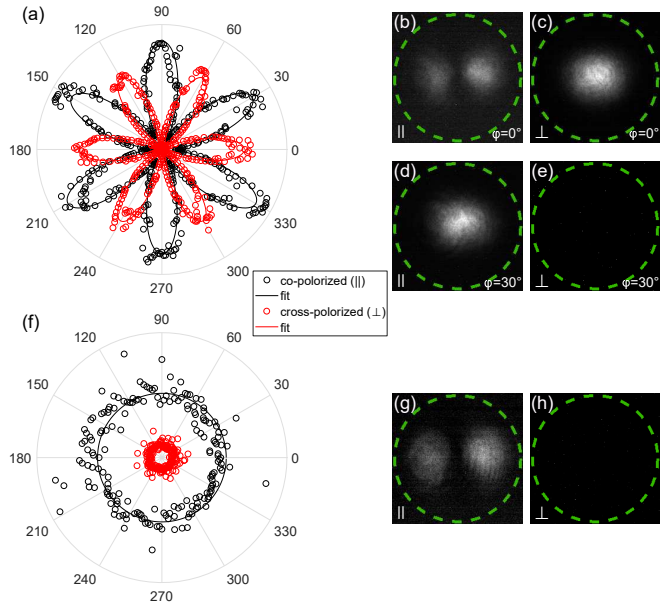}
    \caption{\label{fig3} Polarization-resolved measurements of (a)--(e) mono- and (f)--(h) polycrystalline samples. (a) Polar plot of SH intensity as a function of crystal orientation angle for co-polarized and cross-polarized detection. (b)--(e) Fourier images of the SH emission from the \{111\} surface of a monocrystalline gold flake in co-polarized and cross-polarized detection with orientations $\phi = 0^\circ$ and $\phi = 30^\circ$, respectively. (f) Polar plot, (g) co-polarized, and (h) cross-polarized Fourier images of SH emission from an evaporated gold sample. Green-dashed circles in panels (b)--(e), (g), and (h) indicate the border of the acceptance angle of the objective (NA = 0.9), which also corresponds to the size of the excitation spot in the Fourier plane.}
\end{figure}

As can be seen from Fig.~\ref{fig3}(a), we indeed observe such variation in the SH intensity for the gold flake at 800\,\si{nm} excitation wavelength. By fitting the data to the model we estimate amplitude and phase ratios of the isotropic and anisotropic constants to obtain $a_{i\parallel}/a_{a\parallel} \approx 0.4e^{i91^\circ}$ and $a_{i\perp}/a_{a\perp} \approx 0.1e^{i90^\circ}$. 
In contrast, evaporated gold samples do not exhibit any anisotropy in SHG [Fig.~\ref{fig3}(f)], which can be explained by the fact that the overall response of a polycrystalline film is the average of the individual, randomly oriented domains.

In addition, we study Fourier images of the SH emission, shown in Fig.~\ref{fig3}(b)-(e), (g), and (h). We find that the weak isotropic contribution can be separated from any strong anisotropic contribution when the excitation polarization is parallel to one of the primitive surface lattice vectors $\vec{a}$ (i.e., $\phi=0^\circ$) in co-polarized detection. The depletion of intensity in the center and the petal-shaped intensity distribution indicate that the SH signal is emitted at off-normal angles.
The anisotropic contribution is dominant in cross-polarized detection for the same flake orientation and in co-polarized detection when $\phi=30^\circ$. Its intensity is concentrated mainly in the center of the Fourier images, indicating that SH emission occurs in the direction normal to the sample surface. 
Finally, the polycrystalline gold sample shows only intensity at slanted angles in co-polarized detection, implicitly indicating that only isotropic components of the second-order susceptibility tensor are probed by the out-of-plane electric field components.

\begin{figure}[!ht]
    \centering\includegraphics[width = 0.95\linewidth]{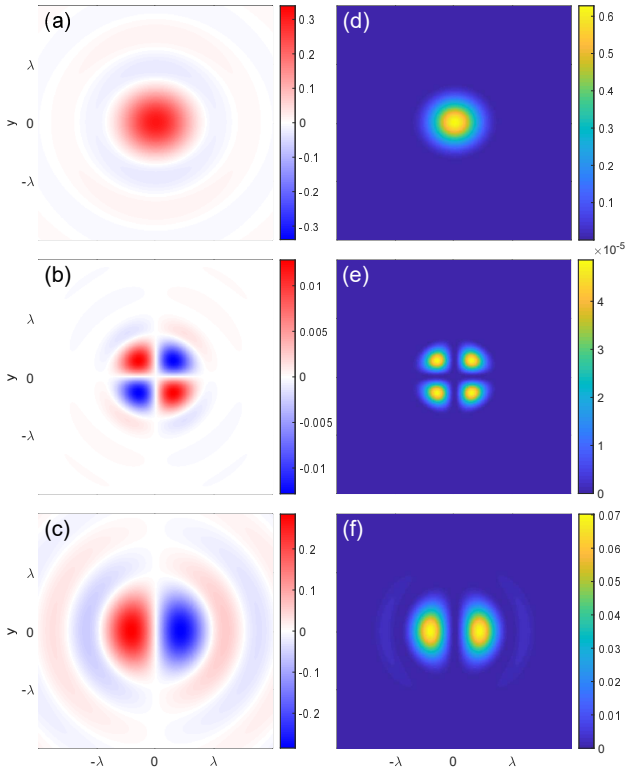}
    \caption{\label{fig4} (a) $E_x$, (b) $E_y$, and (c) $E_z$ components of the electric field at the fundamental frequency $\omega$ in the focal plane of a TFGB. (d) $P_x$, (e) $P_y$, and (f) $P_z$ components of the induced second-order polarization at frequency $2\omega$. The color scales in all plots are normalized to yield the same maximum intensity of the $E$-field at frequency $\omega$.}
\end{figure}

Relative intensities and angular distributions of the isotropic and anisotropic SH emission can be also attributed to the electric field profile of the excitation beam in the focal plane. Fig.~\ref{fig4}(a)-(c) shows the electric field distributions of a tightly focused Gaussian beam (TFGB) calculated by numerical evaluation of the diffraction integral of the angular spectrum, with parameters chosen to approximately match experimental conditions (800\,\si{nm} wavelength, NA = 0.9)~\cite{Principles2012Novotny}. 
Fig.~\ref{fig4}(d)-(f) shows spatial distributions of the second-order polarization induced by the TFGB fields. Clearly, $P_x$, the in-plane component of the polarization parallel to that of the Gaussian beam, is dominant, whereas $P_z$ is one order of magnitude weaker and $P_y$ is nearly absent. Thus, the relative intensity of the anisotropic SH is more prominent, as it is generated only by the strong in-plane field $E_x$, in comparison to the out-of-plane polarization induced by the much weaker $E_z$ component. Additionally, the spatial distribution of $P_z$, which can be interpreted as two dipole-like sources separated by a distance of one wavelength, is consistent with the observed petals in the angular distribution of the isotropic SH [Fig.~\ref{fig4}(d) and (f)], which arises due to far-field interference of the two in-phase oscillating dipoles.

To summarize, we report strong anisotropy in the SHG from monocrystalline gold flakes, which is related to the three-fold symmetry of the $\{111\}$ surface. The anisotropic elements of the  $\chi^{(2)}$ tensor interact mainly with the in-plane electric field components and correspondingly induce in-plane second-order polarization, which leads to SH emission at angles close to the surface normal.
Such anisotropy is absent in polycrystalline gold films due to the random orientation of the crystal domains, which effectively averages out the anisotropic contribution to zero and thus decreases the overall SH emission by approximately one order of magnitude under the given experimental conditions. In contrast, an isotropic SHG component, caused by the out-of-plane fields that produce emission at angles slanted off the surface normal, is present for both poly- and monocrystalline gold surfaces.
We anticipate that our findings will serve as a method for probing sample crystallinity and surface lattice orientation, and may guide future efforts to enhance the efficiency of second-order nonlinear optical processes in high-quality nonlinear plasmonic devices.

\section*{Funding}
C.~W. and T.~Y. acknowledge funding from MULTIPLY fellowships under the Marie Sk\l{}odowska-Curie COFUND Action (grant agreement No. 713694).
A.R.E acknowledges the Secretaria d\text{'}Universitats i Recerca del Departament d\text{'}Empresa i Coneixement de la Generalitat de Catalunya, as well as the European Social Fund (L\text{'}FSE inverteix en el teu futur)-FEDER.
FJGA acknowledges support from the Spanish MINECO (MAT2017-88492-R and SEV2015-0522). 
N.~A.~M. is a VILLUM Investigator supported by VILLUM FONDEN (grant No. 16498) and Independent Research Funding Denmark (grant no. 7026-00117B).
The Center for Nano Optics is financially supported by the University of Southern Denmark (SDU 2020 funding).

\bibliography{AnistropicSHG_arXiv}

\end{document}